\documentclass[12pt,oneside,letterpaper]{article}
\usepackage{amsbsy}
\usepackage{amssymb}
\usepackage{graphicx}
\usepackage[nooneline,bf]{caption}

\newcommand{\bit}{\begin{itemize}}
\newcommand{\eit}{\end{itemize}}
\newcommand{\ben}{\begin{enumerate}}
\newcommand{\een}{\end{enumerate}}
\newcommand{\be}{\begin{equation}}
\newcommand{\ee}{\end{equation}}
\newcommand{\bd}{\begin{displaymath}}
\newcommand{\ed}{\end{displaymath}}
\newcommand{\bea}{\begin{eqnarray}}
\newcommand{\eea}{\end{eqnarray}}
\newcommand{\gapprox}{\;\rlap{\lower 2.5pt\hbox{$\sim$}}\raise 1.5pt\hbox{$>$}\;}
\newcommand{\lapprox}{\;\rlap{\lower 2.5pt\hbox{$\sim$}}\raise 1.5pt\hbox{$<$}\;}

\title{New Problems for a New Century}
\author{R. D. Blandford}
\begin{document} 
\maketitle
\section{Introduction}
This is an optimistic time in X-ray astronomy. Chandra and XMM-Newton
are fully operational and producing results at 
an undigestible rate; Astro-E has been rescheduled for a 2005 launch
and will then assume its place as the third essential component
of contemporary missions. HETE-II is reportedly performing flawlessly
and Swift is on track for a 2003 launch.
Two major international missions, Constellation-X
and XEUS, as well as a host of smaller projects are at various stages
of development. 

My task is to provide some overview of what we have learned
here and, especially, to suggest some questions that should be addressed
in planning new missions. I will have to be somewhat selective and 
idiosyncratic in what I discuss and will refer to the accompanying 
articles, using brackets, for more extensive discussions. As the meeting
has been organised along astronomical lines, I will attempt an orthogonal
organization and try to bring out
some connections by concentrating on the underlying physics.     

However, before doing this, I would like to recall the scientific
context in which Chandra, XMM-Newton and Astro-E were first proposed.
The big issues in 1977, when Chandra was proposed were the X-ray background
and quasars, the two being thought to be closely related. In addition,
there was an understandable envy of optical astronomers and a strong belief
that arcsecond imaging would transform our view of the high energy universe.
For XMM-Newton, in 1982, the prospect of high dispersion spectroscopy 
and the ability to perform abundance analyses and velocity measurements were
strong motivators. By the time that Astro-E was conceived, around 1991, 
the lure of the hard X-ray band and, in particular, iron lines was 
irresistible and it was appreciated that, despite their lower luminosity,
the stronger fluxes from Seyfert galaxies and their more rapid variability 
recommended them as suitable sources with which to study the general 
properties of AGN. In fact, as we have seen here, these have turned out 
to be pretty good scientific justifications for the two missions that
have been launched and the one whose successor we eagerly await. 

X-ray astronomy has become a far richer field than anticipated by these 
proposals. While some problems, like the X-ray background, supernova 
remnants, clusters and accretion disks around cokpact objects appear to 
have been broken, albeit leaving a lot of crucial details to be filled in,
others, like GRBs have become extremely interesting and have developed
in quite unexpected directions. Completely new
phenomena like the Galactic Ridge X-ray Emission and X-ray protostars
of all types, which were widely 
discussed here, have been uncovered. On the technological front there have
been tremendous advances culminating in the microcalorimeter arrays which
are anticipated to lie at the heart of future advances in X-ray 
spectroscopy.   

If I turn now to the future and look at the Constellation-X [White]
brochure I see black hole astrophysics, ``astro-ecology'', and mapping
the distribution of dark matter as the principle scientific challenges.
For XEUS [Jansen], the emphasis is a little more cosmological
and we must add the first groups and quasars, the history of metal 
production and the astrophysics of the hot intergalactic medium. It will
be fascinating to see how these drivers evolve over the coming decade. 
\section{Ten Questions for a New Decade}
In order to enter into the spirit of this meeting, I would like to 
present ten fundamental astrophysical or even physical questions
that I believe can be addressed by X-ray astronomy.
\subsection{How do Cosmic Magnetic Fields really Behave?}
Cosmic magnetic field, on all scales is generally 
described under the MHD approximation. Physically, MHD rests 
on two complementary principles, that magnetic field is frozen 
into moving fluid -- ``go with the flow'' -- and that 
the anisotropic, magnetic stress tensor 
is be responsible for directing the flow  -- ``push-pull''. 
However, even if these 
principles suffice, almost always and everywere, they do not tell the 
whole story. Shocks, reconnection sites and finite Larmor radius effects 
can all lead to behavior that is not consistent with the precepts with the 
precepts of MHD. It is only though observations, interpreted using 
numerical simulation, that we are going to understand how 
magnetized fluid really behaves under cosmic conditions.

X-ray observations reported here are providing vital clues in a
variety of environments. All four types of protostar are found to be X-ray 
sources along with brown dwarfs [Kuboi, Linskey], suggesting that 
coronae are commonly energised, presumably magnetically, to form high 
temperature gas. Disk accretion, likewise, shows signs of magnetic
activity [Mukai, Kubota, Mineshiga]. At the other end of the scale,
many putative black holes in galactic nuclei are found to be spectacularly
underluminous, with powers as low as $\sim10^{-8}L_{{\rm Edd}}$ or 
$\sim10^{-5}L_{{\rm Bondi}}$ [Mushotzky]. This is a very powerful clue as 
to how magnetic viscosity develops in accreting systems. A quite different
consideration is important in rich clusters
of galaxies where the Coulomb scattering 
mean free path is so large that magnetic field must be invoked to limit
the heat transport.
\subsection{How do Cosmic Plasmas really Behave?}
Magnetohydrodynamics is only an approximation to the full, kinetic
behavior of plasma, dealing, as it does, with the large scale conservation
of mass, momentum and energy. Plasma effects may be vital to understanding
the Galactic Ridge emission [Tanaka] observed from the Galactic bulge. 
Collective interactions, like transit time damping,
rather than pure two body Coulomb interactions
may play a role in electron-ion equipartition. These considerations
are of vital importance to understanding stellar coronae [Linsky]
as well as accretion disk coronae and adiabatic accretion in 
the underluminous sources found in galactic nuclei. They are of no less 
importance to understanding the thermal balance of hot gas in clusters and the 
general intergalactic medium and may provide the explanation for why 
lines from intermediate states of ioniziation, such as would be expected
from cooling gas in ionization equilibrium, 
are generally not seen in cluster spectra. 

In addition, I suspect 
that plasma physics will ultimately be needed to resolve the 
controversy concerning the X-ray spectra of Seyfert and LINER accretion 
disks [Kahn, Iwasawa, Brandt, Boller, Nandra, Zdziarski, Reeves, Yaqoob].
However, before this happens some purely observational differences must 
be resolved. How wide and variable are the Fe lines?
How much of the observed spectrum is imprinted by the warm absorber?
Are broad lines of lower Z elements really present in the spectrum?  
\subsection{What is the Structure of Collisionless Shocks?}
Plasma physics is also crucial to the structure of collisionless shocks
which are essential to X-ray astronomy 
as this is almost the only way to heat gas
to high temperature in the interplanetary, interstellar and intergalactic 
media. (Bulk, dissipative heating may be occuring in accretion disks, but
even here, this poses problems.) Shocks also accelerate relativistic electrons.
They inevitably arise when bulk speeds become supersonic either in expanding
flows like stellar winds (Linsky)
or as a consequence of gravitational acceleration
during structure formation in the expanding universe.

Observationally, we are having a hard time locating shocks. There are very few
supernova remnants where we can be sure that we have identified the outer
blast wave [Peter, Aschenbach, Kamae]. Similarly, with the 
accretion shocks around clusters [Bautz, Arnaud]
and even the Galactic center [Koyama]. These shocks are 
responsible for heating gas, accelerating cosmic rays and, possibly, 
stretching magnetic field lines by large factors. However, we 
do not understand electron-ion equipartition, whether cosmic rays 
actually mediate the shock compression and how and whether fields
are amplified. This last point is particularly relevant to the 
mildly relativistic shocks observed in X-ray afterglows from $\gamma$-ray 
bursts, where almost all contemporary models invoke an immediate 
and magical post shock
field enhancement to a fraction of the equipartion value that is just one of 
many free parameters in the fitted spectra [Ricker, Nousek, Fiore]. 
We really need a more basic 
understanding of what is happening and the best approach involves
careful analysis and simulation of the glorious observations of supernova
remnants shown here.
\subsection{When and how were the Elements Made?}
These same observations provide {\it prima facie} evidence that 
heavy elements are synthesized during supernova explosions [Iyudin,
Petre, Aschenbach]. However, the details remain elusive. We really have no
working model of type II supernovae, cannot tell, for example,
whether a neutron star or a black hole has been 
left behind in Cas A and we are in the embarassing 
position that we cannot even agree upon the progenitors of type Ia
supernovae, that are so vital
to contemporary cosmography. X-ray observations really provide an excellent 
handle on the models, probing as they do the mass cut in the explosion of
heavy stars, the extent of the inhomogeneity induced by the explosion,
the strength and angular distribution of the wind 
and the centering which also relates to the recoil speed of the remnant star. 

There are other, conjectured sites of nucleosynthesis, 
especially of r-process
elements, like hypernovae and neutron star coalescence
[Nomoto] and where X-ray observations can be vital, such
as the Fe lines reported from GRBs [Fiore]. On the large scale,
opnservations of distant galaxies can explore abundance gradients and provide
a quantitative measure of the history of star formation [Fabbiano, Iwasawa].
Indeed, claims of abundance enhancements as large as
[Fe/H] $\sim5$ have been made for the centers of galaxies
and in rich clusters [Ohashi, Bohringer].
\subsection{How does Matter Behave at High Density?}
One of the more important ways through which X-ray astronomy can repay 
its debt to fundamental physics is to determine the equation
of state of cold nuclear matter at supranuclear density [Tsuruta, Weisskopf]. 
We really have little idea from basic physics as to whether the interiors
of neutron stars contain free quarks, pion or kaon condensates
or just a proton superconductor plus neutron superfluid
as conservatively! assumed. X-ray observations are the way to find
the answers. This investigation
complements the studies of hot nuclear matter being 
performed at heavy ion colliders and is quite possibly 
relevant to the early universe.

The most direct method is to measure the mass-radius relation of neutron
stars using estimates of the gravitational redshift
and the surface gravity. In practice, this is now 
appearing to be a more difficult task than originally envisaged due to the
absence of spectral lines. As feared, the surface composition is 
largely H and He and there are no strong lines.

A second approach, is less direct but currently more productive. This involves
measuring the cooling rate by comparing the measured surface temperature
with the ages of the surrounding remants. As neutron star atmospheres and 
our understanding of remnant dynamics improve, this should become 
a far more prescriptive as the cooling rates are strongly sensitive to the 
interior composition. 

A third approach is to use the frequencies of neutron star 
QPOs, though this will 
require a much more sophisticated understanding of their dynamics than we have 
at the moment.
\subsection{Is General Relativity Correct?}
In addition to probing nuclear physics, X-ray astronomers have the opportunity
to provide quantitative probes of strong gravity, specifically by detecting
physical effect that are predicted by the relativistic description of the 
spacetime around a spinning black hole [Yaqoob]. The Kerr metric is the 
default description and it will take hard evidence to persuade
most relativists that it is wrong or incomplete. However, it logically 
possible that black holes are more hirsute than conventionally thought 
and that particle (including photon) motion is affected.

To date most attention has centered around attempts to measure the second
parameter, the spin or, equivalently, the specific angular momentum. The width
of the Fe line has been taken as demonstrating that the orbiting gas
gets close to the horizon and that, consequently, holes spin fast.
There are some ongoing concerns about the strength and widths of these lines.
In addition, some ``Devil's Advocates'' have claimed that they can 
be formed by gas plunging into the horizon. However, it will be hard
to make it convincingly quantitative until we understand
the radiative transfer.  I suspect that there is more promise in the 
``diskoseismology'', so comprehensively studied most recently using RXTE
[Swank].
Here, there have also been claims for the detection of spin on the basis 
of some quite simple and to me, not yet compelling descriptions of what is 
going on. However, remarkable patterns are starting to emerge among 
the frequencies and spin must be a major factor in setting the clock.
(The observed QPOs are far too hard for the emission to arise from the 
disk.) It would be wonderful if we could recapitulate the early history 
of atomic spectroscopy and create a redundant, theoretical description 
of these modes that verified the detailed form of the Kerr metric.
\subsection{What is the Nature of Dark Matter?}
An even larger physics challenge is to identify the dark matter and chronicle
its role in the development of large scale structure. Despite the,
perfectly reasonable, focus on neutral supersymmetric particles, and to 
a lesser extent, axions, we really don't understand what it is. 
X-ray astronomy can contribute in a major way by helping trace
the shape of the potential well around ellipticals, 
groups and clusters [Bautz, Boehringer, Ohahshi, Arnaud].
In the last case, velocity dispersions, Sunyaev-Zel'dovich dips and 
weak lensing can all be combined to produce a comprehensive picture.
So far, the most important impression left by the X-ray observations
is that many, though not all, clusters are surprisingly inhomogenous
despite the apparent regularity embodied in the $L-T$ relation.
\subsection{How did the Universe Expand?}
On the larger scale, CMB and SNIa observations have produced an impressive
body of evidence that our 14~Gyr old 
universe is flat, mostly dark and accelerating. Presuming
that this result holds up, it is partly quite unexpected and even more 
perplexing. (It should not be forgotten that, for a long while, the 
most compelling evidence that the Einstein-De Sitter universe was not even 
a close approximation to the truth, came from X-ray observations of clusters
[Henry, Mitsuda].)

The long term kinematic goal must surely be to measure the variation
of the scale factor $a(t)$ with cosmic time in a manner that is independent
of dynamical assumption. (We do not know if the so-called dark energy 
is a manifestation of a shortcoming in the left hand side or the right hand 
side of Einstein's Field Equations, or both.) It is the eternal hope of
observational cosmology that standard rods and candles will be found extending
out to large redshifts 
and X-ray sources, particularly clusters, may yet turn out to be 
examples [Hasinger]. 
\subsection{What were the First Structures in the Universe?}
Although, it is commonly believed that normal galaxies assembled from 
the merger of smaller subunits and that they, in turn congregated
into clusters, we still do not what were the first self-gravitating
structures to form in the expanding universe. What we do know is that 
powerful quasars were already functioning when the universe
was less than $\sim800$ million years old. This implies that 
black holes assembled very early in the life of galaxies 
and probably played an important role in galaxy formation.
Although many of these black holes are likely to be 
obscured, their hard X-ray emission 
can be seen redshifted into the intermediate X-ray energy band.  
   
Alternatively, perhaps the first self-gravitating objects 
were ``Population III''
stars, that formed inside small dark matter halos, 
and which subsequently turned into some of the ``Intermediate Black Holes'' 
the powerful X-ray sources with luminosities in excess of the 
Eddington limit for $\sim100$~M$_\odot$ stars, seen in nearby galaxies
or the X-ray binaries found in globular clusters [Mushotzky].
\subsection{How do you do X-ray Astronomy from the Ground?}
Much of this meeting has been concerned with recent ideas and 
progress in mirror and detector development throughout the entire 
X-ray band [Elvis]. The prospect of pushing focusing optics beyond 
100 keV [Str\"uder, Takahashi] and 
of deploying large microcalorimeter arrays is indeed an exciting one [Stahle]. 
Looking further ahead, the time is surely ripe to start doing  
polarimetry again and work towards making far more sensitive 
X-ray polarimeters. (I note
that it has recent proved possible to measure X-ray circular polarization 
from laboratory synchrotrons.) 

There is even considerable enthusiasm for attempting X-ray interferometry 
from space [Cash]. In its most elemental form this is effectively a 
Young's slit experiment which should appeal, at least in principle, 
to any physicist. How long it takes to develop a system capable of 
accomplishing the stated goal of imaging a black hole remains to be seen,
but there are several important milestones along the way which could
lead to more broadly applicable technology development.  

Among several other approaches to understanding the extreme physical
conditions within the sources are high energy density experiments 
at laser facilities, numerical simulation, especially of three 
(and four) dimensional
magnetohydrodynamics and laboratory astrophysical studies to augment 
our library of wavelengths, oscillator strengths, collision integrals and 
so on that will be necessary for us to interpret X-ray observations 
of hot plasma[Porter].

It will be interesting to see if the questions that have motivated 
proposals for the next generation of X-ray observatories fare as well as
those that motivated Chandra, XMM-Newton and Astro-E.
\section*{Acknowledgements}
I thank Professor Kuneida for the invitation to attend this meeting
and his hospitality and several other participants for helpful advice.
\end{document}